\documentclass{ws-ijmpa}
\usepackage[super,compress]{cite}
\usepackage{graphicx}
\usepackage{color}
\UseRawInputEncoding
\begin{document}
\markboth{V. M. Mostepanenko}{A Few Remarks Concerning Application of the Lifshitz Theory}

%
\catchline{}{}{}{}{}
%

\title{\uppercase{A few remarks concerning application of the Lifshitz theory to calculation
of the Casimir-Polder interaction}}

\author{\uppercase{V. M. Mostepanenko}}

\address{Central Astronomical Observatory at Pulkovo of the
Russian Academy of Sciences, Saint Petersburg,
196140, Russia\\ and\\
Peter the Great Saint Petersburg
Polytechnic University, Saint Petersburg, 195251, Russia\\
vmostepa@gmail.com
}

\maketitle


\begin{abstract}
{The Lifshitz theory} provides a semiclassical description of the Casimir-Polder atom-plate
interaction, where the electromagnetic field is quantized whereas the material of the
plate is considered as a continuous medium. This places certain restrictions on its
application regarding the allowable atom-plate separation distances and the dielectric
properties of the plate material. Below we demonstrate that in some recent literature
the application conditions of the Lifshitz theory established by its founders are violated
by applying it at too short separations and using the dielectric permittivities possessing the
negative imaginary parts in violation of the second law of thermodynamics.
\keywords{Lifshitz theory; dielectric permittivity; second law of thermodynamics.}
\end{abstract}

\ccode{PACS numbers: 12.20.-m, 12.20.Ds, 52.25.Mq}

\newcommand{\ve}{{\varepsilon_{\rm CM}(T_{\Delta},\omega)}}
\newcommand{\rcm}{{\rho_{\rm CM}(T_{\Delta},\omega)}}
\newcommand{\acm}{{a_k^{\rm CM}(T_{\Delta})}}
\newcommand{\wcm}{{\omega_k^{\rm CM}(T_{\Delta})}}
\newcommand{\gcm}{{\gamma_k^{\rm CM}(T_{\Delta})}}
\newcommand{\pgcm}{{{\gamma^{\prime}}_k^{\rm CM}(T_{\Delta})}}
\newcommand{\td}{{(T_{\Delta})}}
\section{Introduction}

The Lifshitz formula for the interaction potential between an atom and a plate expresses
it as a functional of the dynamic polarizability of an atom and the frequency-dependent
dielectric permittivity of the plate material. As established by the founders of the
Lifshitz theory,\cite{1,2,3} it provides a description of the Casimir-Polder interaction
under a condition
\begin{equation}
l \ll z,
\label{eq1}
\end{equation}
\noindent
where $z$ is the atom-plate separation and $l$ is the lattice constant of the plate material.
Just this condition allows to consider the plate material as a continuous medium and use
the idealization of the dielectric permittivity.\cite{1,2,3}

The short-range regime of the Lifshitz formula, where the Casimir-Polder interaction
behaves as $1/z^3$, holds for
\begin{equation}
z \ll \lambda_0,
\label{eq2}
\end{equation}
\noindent
where $\lambda_0$ is the characteristic wavelength for the absorption spectrum of the plate
material.\cite{2,3,4} Calculations show that here much less means that $z$ should be less
than $\lambda_0$ by about a factor of fifty.\cite{5}

As to the long-range regime, where the interaction potential behaves as $1/z^4$, it is
applicable at separations satisfying the conditions\cite{2,3,4}
\begin{equation}
\lambda_0 \ll z \ll \lambda_T,
\label{eq3}
\end{equation}
\noindent
where $\lambda_T \equiv \hbar c/(k_B T)$ is the thermal wavelength, $k_B$ is the Boltzmann
constant, and $T$ is the temperature.

There are also important restrictions imposed on the analytic expressions for the
frequency-dependent dielectric permittivities of plate materials used for calculation of the
Casimir-Polder interaction. Thus, the physically meaningful permittivities must satisfy the
Kramers-Kronig relations and possess the positive imaginary parts. If the latter
requirement is not satisfied, this results in violation of the second law of
thermodynamics.\cite{6}

Below we show that in some recent literature the application regions of the Lifshitz
formula for the Casimir-Polder interaction are replaced with the other ones  in an
unjustified manner, and the proposed analytic expressions for the dielectric permittivity
claiming an excellent accuracy in fact possess the negative imaginary parts over the wide
regions of frequency and temperature.

\section{Application Regions of the Short- and Long-Range Potentials of the Casimir-Polder
Interaction in the Lifshitz Theory}

In place of the application conditions (\ref{eq1}) and (\ref{eq2}) of the short-range potential
established by the founders of the Lifshitz theory, a few papers use the alternative conditions
\begin{equation}
a_0 \ll z \ll \frac{a_0}{\alpha},
\label{eq4}
\end{equation}
\noindent
where $a_0 = \hbar/(m_e c \alpha) = 0.53\,$\AA ~is the Bohr radius and $\alpha$ is the fine
structure constant.\cite{7,8,9}

The conditions (\ref{eq4}), however, are formulated in terms of only the atomic parameters and
disregard the material properties of a plate. For instance, the lattice constant of Si is
$l = 5.45$ \AA. Thus, according to papers,\cite{7,8,9} the short-range regime of the
Lifshitz formula is already applicable at the atom-plate separation $z = l$ because it is
by the order of magnitude larger than the Bohr radius $a_0$. At so short separation,
however, the Si plate cannot be considered as a continuous medium described by the
dielectric permittivity. Thus, in Refs. \citen{7,8,9} the application region of the
Lifshitz theory is incorrectly extended to too short separations.

As to the upper bound of the short-range regime, for Si the characteristic absorption
wavelength $\lambda_0$ is equal to a few hundreds of nanometers. From (\ref{eq2}) we see
that the short-range regime is applicable up to 6--9 nm, but, according to (\ref{eq4}), it
is applicable only at much shorter separations\cite{7,8,9}
$z \ll a_0/\alpha = 7.26$ nm. This means that the upper bound of the short-range regime
is underestimated, as compared to that established by the founders of the Lifshitz theory.

Now we deal with the application conditions (\ref{eq3}) of the long-range Casimir-Polder potential
(note that at $T = 0$ the thermal wavelength $\lambda_T = \infty$). In several
papers,\cite{7,8,9,10} which consider the case of zero temperature, the conditions (\ref{eq3}) are
replaced with
\begin{equation}
7.26~ {\rm nm}=\frac{a_0}{\alpha} \ll z  .
\label{eq5}
\end{equation}

The condition (\ref{eq5}) again disregards the material properties of the plate. In accordance with
(\ref{eq5}), the long-range regime of the Casimir-Polder potential starts at separations exceeding
70 nm, whereas in reality it starts at much larger separations in accordance with the
condition (\ref{eq3}).

\section{General Requirements to the Model Dielectric Permittivities Used in Calculation
of the Casimir-Polder Interaction}

As discussed in Sec. 1, the imaginary part of any physically meaningful dielectric permittivity
must be positive. Recently, the previously considered in the literature Lorentz-Dirac
and Clausius-Mossotti models were used in an ``attempt to find a uniform, simple,
temperature-dependent analytic model for the dielectric permittivity of
monocrystalline (intrinsic) silicon".\cite{8} For this purpose the available experimental
data for the real and imaginary parts of the dielectric permittivity of Si over
the wide frequency and temperature ranges have been fitted to the analytic
expressions suggested by both models. It was claimed that the Clausius-Mossotti
model with two oscillator terms and obtained values of the fitting parameters
reproduces the experimental data for the dielectric permittivity of Si in the
ranges of temperature $293\,\mbox{K}<T<1123\,$K and frequency
$0<\omega<0.16\,\mbox{a.u.}=6.6\times 10^{15}\,$rad/s with an excellent  accuracy.
The resulting permittivity was applied for computation of the Casimir-Polder
interaction potential between a He atom and a Si plate at short and long
separations by means of the Lifshitz theory.

Below it is demonstrated that in the wide ranges of positive frequencies and
temperatures the dielectric permittivity of Si found in Ref.~\refcite{8}
using the Clausius-Mossotti model possesses the negative imaginary part.
This is in contradiction to the fact that dissipation of energy is
accompanied by the emission of heat and, thus, is in violation of the second
law of thermodynamics which is applicable to all bodies in the state of
thermal equilibrium in the absence of electromagnetic field.\cite{6}
Therefore, in these frequency and temperature ranges, the found permittivity
\cite{8} cannot reproduce the valid measurement data with an excellent
accuracy. Thus, the Casimir-Polder energy and other physical quantities
computed using this permittivity are
also under doubt.

In the framework of the Clausius-Mossotti model, the
dielectric permittivity of Si, $\varepsilon_{\rm CM}$, is represented in the
following form\cite{8}:
\begin{equation}
\rcm\equiv\frac{\ve-1}{\ve+2}
 =\sum_{k=1}^{2}
\frac{\acm\left\{\left[\wcm\right]^2-i\pgcm\omega\right\}}{\left[\wcm\right]^2-
\omega^2-i\omega\gcm}.
\label{eq6}
\end{equation}
\noindent
Here, $\wcm$ are the resonance frequencies, $\gcm$ are the level widths,
$\pgcm$ are the radiation damping constants, and $\acm$ are the amplitudes.
All these fitting parameters depend on the
temperature $T$ via the dimensionless quantity
$T_{\Delta}=(T-T_0)/T_0$,
where $T_0=293\,$K. Thus, $T_{\Delta}$ varies from 0 to 2.833. This corresponds
to the range of $T$ from 293\,K to 1123\,K.

According to Ref.~\refcite{8}, the dielectric permittivity $\ve$ satisfies the
Kramers-Kronig relations and its real and imaginary parts are the even and odd
functions of frequency, as it should be for any function which claims to play
the role of dielectric permittivity. However, it does not satisfy the condition
\begin{equation}
{\rm Im}\varepsilon(T_{\Delta},\omega)>0,
\label{eq8}
\end{equation}
\noindent
which must be valid for all bodies in the state of thermal equilibrium with the
environment in the absence of alternating electromagnetic field in accordance
with the law of entropy increase (the second law of thermodynamics).\cite{6}.
Really, from (\ref{eq6})
one easily obtains
\begin{equation}
{\rm Im}\ve=\frac{3{\rm Im}\rcm}{[1-{\rm Re}\rcm]^2+[{\rm Im}\rcm]^2},
\label{eq9}
\end{equation}
\noindent
where
\begin{equation}
{\rm Im}\rcm=\!\sum_{k=1}^{2}\!\acm\omega
\frac{[\wcm]^2
[\gcm-\pgcm]+\omega^2\pgcm}{\left\{\left[\wcm\right]^2
-\omega^2\right\}^2+\omega^2\left[\gcm\right]^2}.
\label{eq10}
\end{equation}

{}From (\ref{eq9}) it is seen that the sign of ${\rm Im}\varepsilon_{\rm CM}$
coincides with  the sign of ${\rm Im}\rho_{\rm CM}$. Next, from (\ref{eq10})
one concludes that  ${\rm Im}\rho_{\rm CM}$ and, thus, ${\rm Im}\varepsilon_{\rm CM}$
are negative if the following condition is satisfied for both $k=1$ and $k=2$:
\begin{equation}
\omega<\wcm\sqrt{1-\frac{\gcm}{\pgcm}}.
\label{eq11}
\end{equation}

By using the values of the fitting parameters presented in Tables I and II of
Ref.~\refcite{1}, one finds that the inequality (\ref{eq11}) is satisfied
for both $k=1$ and $k=2$ at $T_{\Delta}=0.614$, 0.785, 0.956, 1.126. 1.397,
and 1.468. At the corresponding temperatures $T=472.9\,$K, 523.0\,K, 573.1\,K,
622.9\,K, 702.3\,K and 723.1\,K the imaginary part of $\ve$ takes the negative
values over the frequency ranges from 0 to $5.3\times 10^{14}$, $8.2\times 10^{14}$,
$1.33\times 10^{15}$, $1.36\times 10^{15}$, $1.47\times 10^{15}$, and
$1.62\times 10^{15}\,$rad/s, respectively. All these frequency ranges belong to
the range from 0 to $0.16\,\mbox{a.u.}=6.6\times 10^{15}\,$rad/s where
an excellent accuracy of the dielectric function of Si obtained
using the Clausius-Mossotti model is claimed.\cite{8} As an example, the negative
imaginary part of $\ve$ at $T=573.1$~K is shown in Fig. 1.
\begin{figure}[b]
\vspace*{-13.2cm}
\hspace*{-2.5mm}
\centerline{\includegraphics[width=15.5cm]{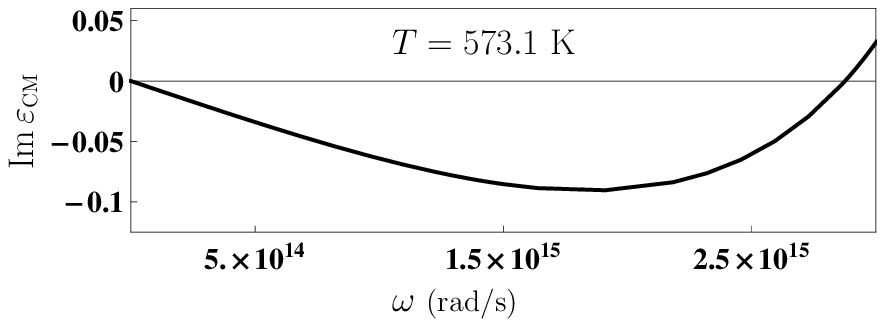}}
\vspace*{-5.5cm}
\caption{The imaginary part of $\varepsilon_{\rm CM}$ at low frequencies at $T=573.1$~K.
 \label{f1}}
\end{figure}

An approximation of the fitting parameters by quadratic functions made in Eq.~(11)
of Ref.~\refcite{8} does not remedy this defect. By using the coefficients of
quadratic functions presented in Table~III of Ref.~\refcite{8}, one finds that
${\rm Im}\ve$ remains negative in the range of $T_{\Delta}$ from 0.375 to 1.469,
i.e., from approximately $T=403\,$K to 723\,K. This is even a wider temperature
region than that obtained directly from the fitting parameters of Tables~I
and II do not using any approximation.

One additional remark concerning the dielectric permittivity of Si obtained\cite{8}
by using the Lorentz-Dirac model
\begin{equation}
\varepsilon_{\rm LD}(T_{\Delta},\omega)=1+\sum_{k=1}^{2}\frac{a_k^{\rm LD}\td
\left\{\left[\omega_k^{\rm LD}\td\right]^2 -i{\gamma^{\prime}}_k^{\rm LD}\td
\omega\right\}}{\left[\omega_k^{\rm LD}\td\right]^2-\omega^2-
i\omega{\gamma}_k^{\rm LD}\td}
\label{eq12}
\end{equation}
\noindent
is in order. Similar to the above analysis, we see that the term
of $\varepsilon_{\rm LD}$ with  $k=1$ computed with the fitting parameters
defined in Table IV of Ref.~\refcite{8} possesses the negative imaginary part at
all values of $T_{\Delta}$ from 0 to 2.833. Based on the laws of thermodynamics,
one concludes that such a result contradicts to the physical meaning of this term
as describing the first absorption peak of monocrystalline Si.

The obtained dielectric permittivities were used to calculate the coefficients
$C_3$ and $C_4$ in the short-range, $C_3\td/z^3$, and long-range, $C_4\td/z^4$,
asymptotic behavior of the interaction potential between a Si plate and a He
atom spaced at the height $z$ above it. Calculations were performed by means of
the Lifshitz theory of atom--plate interaction using the dielectric permittivity of Si
along the imaginary frequency axis. The obtained results cannot be considered
as fully reliable even if to admit that the frequency regions, where the relatively
small in magnitude imaginary part of $\varepsilon_{\rm CM}$ is negative, make
rather small impact on $\varepsilon_{\rm CM}(i\omega)$. It should be also kept
in mind that the ``excellent" analytic expressions for the dielectric
permittivity of Si may be used not only for
calculations of the atom-wall potentials, but in studying diverse physical
phenomena fully determined by the behavior of this permittivity at relatively low
real frequencies, where the suggested expressions are rudely wrong. One can mention
the Casimir and Casimir-Polder forces out of thermal equilibrium, the radiative heat
transfer, the near-field spectroscopy, etc.

\section{Conclusions and Discussion}

To conclude, computations of the Casimir-Polder interaction using the Lifshitz theory
are sometimes made outside the region of its applicability and, specifically, at too
short atom-plate separations, where the plate material cannot be considered as a
continuous medium. To justify such an approch, the physically well grounded
application regions of both the short- and long-range Casimir-Polder potentials
established by the founders of the Lifshitz theory, are revised by disregarding the
atomic stricture of a plate material.

The claimed ``excellent accuracy'' of the analytic expression for the dielectric
permittivity of Si used in computations of the Casimir-Polder interaction is incorrect
because the imaginary part of this permittivity is negative over the wide frequency and
temperature ranges in violation of the second law of thermodynamics.

Finally, computations of the van der Waals (Casimir-Polder) interaction at separations
below a few nanometers should be performed not by means of the Lifshitz theory
but, e.g., by the methods of molecular dynamics accounting for the atomic structure
of a plate material.

\section*{Acknowledgments}

This work was supported by the
State Assignment for basic research (project FSEG--2023--0016).

\end{document}